\documentstyle[12pt,epsfig]{article}

\def\be{\begin{equation}}
\def\ee{\end{equation}}
\def\ba{\begin{eqnarray}}
\def\ea{\end{eqnarray}}
\def\ga{\mathrel{\raise.3ex\hbox{$>$\kern-.75em\lower1ex\hbox{$\sim$}}}}
\def\la{\mathrel{\raise.3ex\hbox{$<$\kern-.75em\lower1ex\hbox{$\sim$}}}}
\def\any{\mathrel{\raise.3ex\hbox{$<$\kern-.75em\lower1ex\hbox{$>$}}}}
\newcommand{\sect}[1]{\section{#1}\setcounter{equation}{0}}

\newcommand{\beq}{\begin{equation}}
\newcommand{\eeq}{\end{equation}}
\def\beqa{\begin{eqnarray}}
\def\eeqa{\end{eqnarray}}
\def\p{\partial}

\def\nn{\nonumber}
\def\q{\quad}
\def\t{\tilde}
\def\tn{\tilde{\nabla}}
\def\mn{{\mu \nu}}
\def\bc{\begin{center}}
\def\ec{\end{center}}

\def\vf{\varphi}
\def\th{\theta}

\begin{document}
\begin{titlepage}
\rightline{CERN-TH/2002-162}
\rightline{hep-th/0207283}
\rightline{July 2002}
\begin{center}

\vspace{1cm}

\large {\bf Schwarzschild Black Branes and Strings\\
 in higher-dimensional Brane Worlds}\\
\vspace*{5mm}
\normalsize

{\bf  P. Kanti$^1$},  {\bf  I. Olasagasti$^2$} and
{\bf K. Tamvakis$^2$}

\smallskip
\medskip
$^1${\it CERN, Theory Division,\\
CH-1211 Geneva 23, Switzerland}

 \vspace*{3mm}

$^2${\it Physics Department, University of Ioannina,\\
GR-451 10 Ioannina, Greece}

\smallskip
\end{center}
\vskip0.6in

\centerline{\large\bf Abstract}

We consider branes embedded in spacetimes of codimension one and two, with a
warped metric tensor for the subspace parallel to the brane. We study
a variety of brane-world solutions arising by introducing a Schwarzschild-like
black hole metric on the brane and we investigate the properties
of the corresponding higher-dimensional spacetime. We demonstrate that normalizable bulk modes lead to a vanishing flow of
energy through the naked singularities. From this point of view, these singularities are harmless.

%%%%%%%%%%%%%%%%%%%%%%%%%%%%%%%%%%%%%%%%%%%%%%%%%%%%%%%%%%%%%%%%%%%%%
%\vfill
%\vskip 0.15in
%\leftline{CERN--TH/2001-109}
%\leftline{April 2001}
\end{titlepage}
%\baselineskip=18pt
%%%%%%%%%%%%%%%%%%%%%%%%%%%%%%%%%%%%%%%%%%%%%%%%%%%%%%%%%%%%%%%%%%%%%

\sect{Introduction}

Higher dimensional models of Gravitation are motivated by the need to explain
the large difference in magnitude between two important energy scales in
nature: the Planck scale $M_P\sim 2\times 10^{18}$ GeV that governs the
gravitational interactions, and the Electroweak scale of particle physics.
The study of these models during the last few years has led to a number of
interesting theoretical ideas \cite{large1,RS1,RS2}, among them the
possibility of extra non-compact dimensions \cite{RS2} and the effective
localization of gravity. In the latter case, the higher-dimensional spacetime
is filled with a negative cosmological constant that gives rise to an
$AdS$ spacetime. The Standard Model interactions are confined on zero-thickness
{\textit{$3$-branes}}, while the graviton can propagate in the five-dimensional
{\textit{bulk}}. Four-dimensional gravity is successfully reproduced on the
brane by an ordinary massless graviton localized on it \cite{graviton}, while
the four-dimensional Planck mass is merely an effective scale derived from
the fundamental energy scale and the $AdS$ radius of the five-dimensional
spacetime.

In a five-dimensional framework, as the Randall--Sundrum model, it would be
natural to expect that when matter, trapped on the brane, undergoes
gravitational collapse, the four-dimensional background that forms would be
described by a Schwarzschild-type metric. The first attempt to substitute
the flat, Minkowski background on the brane by a Schwarzschild spacetime,
in the context of the one-brane RS model, led to the derivation of the
RS--Schwarzschild black hole solution ~\cite{CHR}. The induced metric on
the brane comes out to be purely Schwarzschild, as expected, respecting
all the usual astrophysical constraints~\cite{Giannakis}. However, the
solution, when seen from the five-dimensional point of view, describes
a {\textit{black string}}, an extended black-hole singularity, infinite
in the fifth dimension. As the study of the curvature invariant quantities
of the theory reveal, the AdS horizon, located at infinity, has also been
changed to a true, naked singularity of the five-dimensional spacetime.
A localized {\textit{black cigar}} solution, with a finite extension along
the fifth dimension, might arise due to a Gregory-Laflamme type of
instability~\cite{GL} that appears near the $AdS$ horizon.

The subject of black holes in the context of extra dimensions has been
extensively studied during the last few years \cite{bh-brane}-\cite{large}.
In a recent paper \cite{KT}, two of us considered four-dimensional metric
ansatze of the Schwarzschild type with a horizon depending on the fifth
dimension aiming at finding localized black hole solutions. Nevertheless,
such configurations turned out to be compatible only with exotic bulk-matter
distributions. Although in these solutions the harmless character of the
AdS horizon was restored and the black hole singularity rapidly `decayed'
away from the brane, the model was infested with an additional singularity
located at the induced horizon of the black hole.

It has become clear by now that the embedding of a four-dimensional
black-hole into a higher-dimensional spacetime leads to the change of the
topological properties of the $d$-dimensional spacetime in a dramatic way:
the four-dimensional black-hole singularity becomes manifest also in the
higher-dimensional curvature invariants, while previously harmless
$d$-dimensional horizons now become true spacetime singularities. In
the present article, we consider models of maximum codimension $2$, with
a factorizable metric for the subspace parallel to the brane. We perform
a singularity analysis, that tries to shed light on the topological structure
of various five-, six-, or higher-dimensional spacetimes of the particular
type, and we examine the inevitable appearance of black-hole and naked
singularities in the square of the Riemann tensor. Whenever necessary, the
naked singularity is shielded by a second brane, which automatically leads
to a finite-size {\it black tube} construction that interpolates between two
four-dimensional Schwarzschild-like black holes. In solutions, however, with
a cosh-like profile for the warp factor, along one or more extra dimensions,
an {\it effective localization} of both the black-string and the bulk singularity
may take place through the suppression of the singular terms, appearing
in the curvature invariants, away for the brane. In the context of
single-brane configurations, the bulk singularity may terminate the
infinitely-extended black-string, however, a non-vanishing flow of energy
through the naked singularities of the spacetime would lead to the breakdown
of the conservation of currents associated with spacetime symmetries. We
therefore examine whether normalizable bulk modes lead to a vanishing
flow of energy and render the naked singularities harmless from this
point of view.

The outline of this article is as follows: Section 2 presents our
spacetime ansatz and useful formulae to be used in the following analysis.
In Section 3, we focus on the study of 5-dimensional brane-world solutions
and we consider three categories depending on the profile of the warp
factor along the extra dimension. Section 3 extends our analysis in
higher-dimensional models and we study a variant of the Cohen-Kaplan
\cite{CK} global string model, which has cylindrical symmetry in the
transverse space, and a $d$-dimensional version of the cylindrically
non-symmetric model of Kanti, Madden and Olive \cite{Kanti:2001vb} (KMO).
We present our conclusions in Section 5.

%%%%%%%%%%%%%%%%%%%%%%%%%%%%%%%%%%%%%%%%%%%%%%%%%%%%%%%%%%%%%%%%%%%%%%%%%%%

\section{Models of General Codimension}

In what follows
\footnote{We consider the standard Einstein gravitational Action
$${\cal{S}}=\int d^dX\sqrt{-G}\left\{2M^{d-2}R-\Lambda_B\right\}
-\int d^{d}X\sqrt{-G}\,{\cal{L}}
-\int d^{(d-1)}x\sqrt{- h}\,\hat{\cal{L}}\,,$$
where the last term corresponds to a localized source on a wall
with induced metric $h_{a b}$.}
%The addition of a brane at a fixed location amounts to a term
%$-\sigma \int d^4x\,\sqrt{-\hat{g}}$.}
we are going to consider warped geometries of codimension
higher than $1$. The total number of dimensions $d$ is the sum of $q$
{\textit{longitudinal}} dimensions along the brane (occasionally we will
specialize in the case $q=4$) and $n$ {\textit{transverse}} dimensions.
By warped geometries we mean all those with line-elements that
can be written as
\beq
ds^2=
e^{2A(y)} d\hat{s}_q^2+d\tilde{s}_n^2\equiv
e^{2A(y)}\hat{g}_\mn (x)\,dx^{\mu} dx^{\nu}+\gamma_{ab}(y)\,dy^a dy^b\,.
\label{warp}
 \eeq
In the above metric, $d\hat{s}^2$ is a $q$-dimensional line-element that
depends only on the longitudinal coordinates $\{x^{\mu}\}$. Likewise,
$d\tilde{s}_n^2$ refers to an $n$-dimensional line-element that depends
only on the transverse coordinates $\{y^{a}\}$. Henceforth, we will be
using a hat and a tilde in order to denote quantities evaluated exclusively
in terms of the $\hat g_\mn$ and $\gamma_{ab}$ metric tensors, respectively.

For the above general warped metric, the various components of the Ricci
tensor are
%%%%%%%%%
\beqa
R_{a b}&=&\tilde{R}_{a b}-q\,({\tilde{\nabla}_a \tilde{\nabla}_b A
+\tilde{\nabla}_a A \tilde{\nabla}_b A})\,,\\[1mm]
R_\mn &=& \hat{R}_\mn - \hat{g}_\mn \;e^{2A}
[ {\tilde {\nabla}}^2 A+q\,({\tilde{\nabla} A})^2 ]\,,\\[1mm]
R_{a \mu}&=&0\,.\eeqa
The resulting Ricci scalar is
\beqa
R=\tilde{R}_{(n)}+{\hat{R}}_{(q)}\,e^{-2A}-2q ({\tilde {\nabla}}^2 A)
-q(q+1)(\tilde{\nabla}A)^2\,.
\label{A}
\eeqa

Another quantity of special interest is the square of the Riemann
tensor\footnote{The Riemann tensor components different from zero are
${\tilde R}^{a}_{bcd}$ and
$$R^{\alpha}_{\; b \nu d}=-\delta^{\alpha}_{\nu}(\tn_b \tn_d A+\tn_b A \tn_d A)
\,,\,\,\,\,\,\,\,\,R^a_{\; \mn d}=g_{\mn} (\tn_d \tn^a A+\tn^a A \tn_d A)\,,$$
$$R^{\alpha}_{\; \mn \lambda}= \hat{R}^{\alpha}_{\; \mn \lambda}+
(\tn A)^2(g_{\mn}\delta^{\alpha}_{\lambda}-
g_{\mu \lambda}\delta^{\alpha}_{\nu})\,.$$},
that reveals, more accurately than any other scalar quantity, the
singular properties of spacetimes containing black-hole line-elements.
It can be computed to be
\beqa
(R^A_{\;BCD})^2&=& (\tilde{R}^a_{\; bcd})^2+e^{-4A}(\hat{R}^{\alpha}_{\;
\mn \lambda})^2-4e^{-2A}(\tn A)^2 \hat{R}_{(q)} \nn \\[1mm]
&+& 4q\,(\tn_a \tn_b A + \tn_a A \tn_b A)^2 + 2q(q-1)(\tn A)^2 (\tn A)^2\,.
\label{B}
\eeqa

%%%%%%%%%%%%%%  mych1
From the Einstein tensor for the warped metrics,
we can read out the structure of energy-momentum tensors compatible with it,
%%%%%%%%%%%%%  end mych1
\beqa
T_{ab}&=&\t T_{ab}(y)-{1\over 2}\,\gamma_{a b}(y)\,e^{-2A(y)}
{\hat R}_{(q)}(x)\,, \\
T_\mn&=&\hat T_\mn(x)+\hat g_\mn(x)\,\Delta(y)\,,\\[1mm]
T_{a \mu}&=&0\,.
\eeqa
$T_{ab}$ depends on $\{x^{\mu}\}$ only through $\hat R_{(q)}(x)$,
the Ricci scalar of $d \hat s_q^2$. $T_\mn$, on the other hand,
depends on $\{y^{a}\}$ only through a term proportional
to $\hat g_\mn(x)$, $\Delta(y)$, which is not universal but
depends on the particular source in the bulk.

For metrics such as (\ref{warp}), the
equation for $\hat{g}_\mn$ decouples. In fact, one can show
that the Einstein tensor for $\hat{g}_\mn$ is
\beq
\hat{G}_\mn =\hat T_\mn - K \hat g_\mn\,,
\eeq
where $\hat T_\mn$ is the $x^{\mu}$-dependent part of the energy-momentum
tensor and what appears here as a cosmological constant term $K$
is just a separation constant. This means that as long as we fix
$K$, independently of the explicit form of the
$4D$ solution, the geometry along the transverse space and
the warp factor will remain the same. This is true in particular for
all the solutions with $\hat T_\mn=0$. There
are many examples of this class in the literature \cite{Ola-1}.
For the most part, the existing
solutions consider the maximally symmetric $4D$ solutions.
It is clear however from the discussion above, that we can generalize
any of them by considering  the corresponding black hole solution
counterparts. This will work both for Schwarzschild and Kerr
black holes but not for charged ones since the energy-momentum
tensor for the charged black hole on the brane does not
fall into the class described by the above warped geometries \cite{Ola-2}.

An important thing to note is that, for any non-trivial $4D$
geometry, any zero of the warp factor will give rise to a naked singularity.
We can see this from the expressions (\ref{A}) and (\ref{B}). If either
$(\hat{R}^{\alpha}_{\; \mn \lambda})^2$ or $\hat{R}_{(q)}$ is non-zero
due to a non-trivial ${x^{\mu}}$-dependence, a singularity will
develop wherever the warp factor goes to zero.

A naked singularity could be tolerated and considered harmless if no
conserved quantities flow into the singularity. If $K^A$ is a Killing vector,
this will give rise to an associated conserved current $J^A=T^A_B K^B$.
If the horizon has a normal vector $n^A$ and we define $z$ locally by
$dz=n^A dX_A$, then the flow of the conserved current into the singularity,
at $z=0$, is given by
\beq\mathop{\lim}\limits_{z\to 0}\;\;
 \int_{z} n_A J^A \; d\sigma,
 \eeq
where $d\sigma$ is the invariant volume element
of the constant $z$ hypersurfaces.
So, the condition for vanishing flow is
\beq
\mathop{\lim}\limits_{z\to 0}\;\;
 n_A J^A\;
d\sigma=
\mathop{\lim}\limits_{z\to 0}\;\;
 n_A T^A_B K^B\;
d\sigma
=0\,.
\eeq
For a scalar field $\phi(x,y)$ propagating in the bulk, the condition of
vanishing flow reduces to
\beq
\mathop{\lim}\limits_{z\to 0}\;\;\p_z \phi \; K^A\p_A \phi \;
d\sigma=0\,.
\label{flow}
\eeq
In the following sections, we will apply the above scheme on five-, six,
or higher-dimensional models and look for scalar modes that respect the
above constraint and ensure vanishing flow of conserved currents through
the naked singularities of the spacetime.

%%%%%%%%%%%%%%%%%%%%%%%%%%%%%%%%%%%%%%%%%%%%%%%%%%%%%%%%%%%%%%%%%%%%%%%%

\sect{The 5-dimensional Case}

In this section, we focus on the case of a five-dimensional spacetime,
whose line-element may be written in the factorized form
%%%%%%%%%%%%%
\be
ds^2=g_{MN}\,dx^M\,dx^N =
e^{2A(y)}\hat{g}_{\mu\nu}(x)dx^{\mu}dx^{\nu}+dy^2\,.
\label{5-fact}
\ee
%%%%%%%%%%%%%
The four-dimensional metric, $\hat g_{\mu\nu}$, is assumed to describe a
spherically symmetric black-hole background. As mentioned in the introduction,
we expect that the embedding of the four-dimensional black-hole into the
five-dimensional spacetime will lead to the manifestation of the 4D
black-hole singularity in the five-dimensional curvature invariants.
Moreover, five-dimensional horizons located at the boundaries of the
extra dimension will now become true spacetime singularities. In this
section, we perform, first, a singularity analysis that tries to shed
light on the topological structure of various five-dimensional spacetimes
of the type (\ref{5-fact}), and secondly, a study of the possible flow
of energy through the true singularities of the spacetime.

Starting from the ansatz (\ref{5-fact}) we now consider three different
cases of five-dimensional spacetimes, defined by the behaviour of the
warp factor $e^{A(y)}$ along the extra dimension.

\bigskip

{\textbf{1. Asymptotically vanishing warp factor.}}

\medskip

In this category belongs the class of line-elements that are characterized
by the vanishing of the warp factor $e^{A(y)}$ at an infinite
coordinate distance from the brane. A typical example of this type is the
spacetime describing the five-dimensional {\it black string} solution
\cite{CHR} with line-element
%%%%%%%%
\be
ds^2=e^{-2 k |y|}\,\biggl\{-\biggl(1-\frac{2M}{r}\biggr)\,dt^2 +
\biggl(1-\frac{2M}{r}\biggr)^{-1}\,dr^2 + r^2(d\theta^2 +
\sin^2\theta\,d\varphi^2) \biggl\} + dy^2\,,
\label{metric-bs}
\ee
%%%%%%%%%%%
where $k^2=\kappa^2_5 |\Lambda_B|/6$. In this case, the five-dimensional
bulk is filled with a negative cosmological constant $\Lambda_B$, and the
line-element remains invariant under the mirror transformation
$y \leftrightarrow -y$, with the infinitely-thin, positive-tension brane
located at $y=0$. By making use of Eq. (\ref{B}), the Riemann
curvature invariant for the above background comes out to be
%%%%%%%%%%%%%
\be
R_{ABCD}\,R^{ABCD}= 40 k^4 + \frac{48 M^2 e^{4 k |y|}}{r^6}\,,
\label{Riemann-RS}
\ee
%%%%%%%%%%%%%
and it clearly diverges at the black-hole singularity at $r=0$, and the
AdS horizon $|y| \rightarrow \infty$. For factorized four-dimensional
line-elements like the one above, the only method, that makes the AdS
singularity disappear and gives a finite size to the black-hole singularity,
is the introduction of a second,  negative-tension brane at $y=L < \infty$.
The induced metric tensor on both branes will describe a four-dimensional
black hole. If we choose the inter-brane distance to be the one that resolves
the hierarchy problem \cite{RS1}, then, a {\it black tube} construction,
extending into the bulk, will interpolate between two brane black holes
whose masses, and horizon values, will differ by 16 orders of magnitude.

It is nevertheless interesting to investigate whether
we can have
a non-vanishing flow
of energy into the bulk singularities
in the case of the
single-brane configurations, when the second brane is absent. Here,
we follow the method of Ref. \cite{CK},  however, our analysis considerably
deviates from theirs at certain points due to the different metric ansatz.
The equation of a massless scalar field propagating in the background
(\ref{5-fact}) may be written as
%%%%%%%%%%%%%
\be
e^{-2A}\,\partial_y\,\Bigl(e^{4A}\,\partial_y \Phi_y\Bigr) +
m^2 \Phi_y=0\,,
\label{eq1}
\ee
%%%%%%%%%%%%%%
under the assumption that the scalar field has the factorized form
$\Phi(x,y)=\Phi_x(x)\,\Phi_y(y)$, and
%%%%%%%%%%%%%%%
\be
\frac{1}{\sqrt{-\hat g}}\,\partial_\mu\,\Bigl[\sqrt{-\hat g}\,
\hat g^{\mu\nu}\,\partial_\nu \Phi_x \Bigr]=m^2 \Phi_x\,.
\label{eq-4D}
\ee
%%%%%%%%%%%%%%%

We first look for solutions to the above equation (\ref{eq1}) describing
massless modes, $m^2=0$. Then, Eq. (\ref{eq1}) clearly reveals the existence
of two zero-mode solutions:
%%%%%%%%%%%
\be
\Phi^{(1)}_y=const.\,, \qquad  \quad
\Phi^{(2)}_y= const. \,\int^y \,\frac{dy'}{e^{4 A(y')}}\,.
\label{sol}
\ee
%%%%%%%%%%%
The metric ansatz (\ref{metric-bs}) is characterized by translation
invariance along the $(t,\theta,\varphi)$ coordinates and rotational
invariance in the 4-dimensional spacetime. For our purposes, it is
enough to consider the Killing vector $K_M^{(\mu)}=e^{2 A}\,\delta_M^\mu$,
with $\mu=t,\theta,\varphi$, and demand that no flow of energy,
associated with the current $J^y_{(\mu)}=T^{y N} K_N^{(\mu)}$, takes
place at the location of the bulk singularity. That means we should have
%%%%%%%%%%%%%
\be
\lim_{|y| \to \infty} \sqrt{-g}\,J^y_{(\mu)}=
\lim_{|y| \to \infty} \,\Bigl[\sqrt{-\hat g}\,\hat g^{\mu\nu}\,
e^{4A}\,\Phi_x\,\bigl(\partial_x \Phi_x\bigr)\,
\Phi_y\,\bigl(\partial_y \Phi_y\bigr)\Bigr]=0\,.
\label{con1}
\ee
%%%%%%%%%%%%%
Taking into account the behaviour of the two solutions (\ref{sol})
near the AdS horizon, $\Phi^{(1)}_y=const.$ and $\Phi^{(2)}_y=const.
\,e^{4 k|y|}$, we may see that the flow of energy is indeed zero for
the first solution while diverges for the second one. We need,
therefore, to discard the latter as an {\it irregular} solution
and keep only the former as the physically acceptable, {\it regular}
solution. As a result, for the massless modes, it is only the
scalar configurations with a trivial profile along the extra
dimension that are acceptable.

We may easily check that, under the assumption that $\Phi_y=const.$, all
the remaining components of the current $J^M$ also vanish at the AdS boundary.
We should stress at this point that the four-dimensional line-element has
a non-trivial topology that introduces an extra singularity, the black-hole
one, in the five-dimensional spacetime. This singularity is protected
by a horizon and thus it is harmless. As for all scalar modes propagating
in a black-hole background, we demand that they are well defined in the
regime $r > r_H=2M$ and have a finite value at the black hole horizon.

We now turn to the massive scalar modes. By performing a coordinate
transformation $dy=e^A\,dz$, and going to conformal coordinates,
Eq. (\ref{eq1}) may be written as
%%%%%%%%%%%%%
\be
\biggl[ -\frac{d^2 \,}{dz^2} + \frac{\Psi''}{\Psi} \biggr]
\,\phi(z)=m^2 \phi(z)\,,
\label{eq2}
\ee
%%%%%%%%%%%%%%
under the assumption that $\Phi_z(z)=\phi(z)/\Psi(z)$, where
$\Psi(z)=e^{3A(z)/2}$. In terms of the new coordinates,
$e^{A(z)}=(k\,|z + z_0|)^{-1}$ and therefore, the
term $\Psi''/\Psi$ behaves, near the AdS boundary, like
$1/z^2$ and vanishes as $z \rightarrow \infty$. As a result,
this term is sub-dominant compared to the mass term near the
singularity, a fact which, when taken into account, leads to
the following solution for the $z$-dependent part of the scalar
field
%%%%%%%%%%%%%%%
\be
\Phi_z(z)=(k z)^{3/2}\,\Bigl[C_1\,e^{i m z} + C_2\,e^{-i m z}\Bigr]\,,
\label{sol-m}
\ee
%%%%%%%%%%%%%%%
where $C_1$ and $C_2$ are integration constants. The scalar modes in this
case are given by plane waves with a $z$-dependent amplitude and lead
indeed to a vanishing flow of the fifth component of the conserved current
through the boundary of the extra dimension. This may be seen by using the
constraint  (\ref{con1}), written in conformal coordinates
%%%%%%%%%%%%%
\be
\lim_{|z| \to \infty} \,\Bigl[\sqrt{-\hat g}\,\hat g^{\mu\nu}\,
e^{3A}\,\Phi_x\,\bigl(\partial_x \Phi_x\bigr)\,
\Phi_z\,\bigl(\partial_z \Phi_z\bigr)\Bigr]=0\,,
\ee
%%%%%%%%%%%%%
and substituting the solution (\ref{sol-m}). Then, the {\it lhs} of the above
expression turns out to be proportional to a wildly oscillating function with
unit amplitude, that reduces to zero.

\bigskip

{\textbf{2. Vanishing warp factor at finite coordinate distance.}}

\medskip

Another particular type of solutions, that arises as a special case of
the metric ansatz (\ref{5-fact}), is the one that is characterized by a
vanishing warp factor at finite coordinate distance, as opposed to
infinite distance in the previous category. A particular example is
the RS--dS--Schwarzschild black hole \cite{Kaloper,Kim,KR,HK} with
line-element
%%%%%%%%
\be
ds^2=e^{2 A(y)}\,\biggl\{-\biggl(1-\frac{2M}{r}- \Lambda r^2\biggr)\,dt^2 +
\biggl(1-\frac{2M}{r}- \Lambda r^2\biggr)^{-1}\,dr^2 + r^2\,d\Omega^2
%(d\theta^2 +
%\sin^2\theta\,d\varphi^2)
\biggl\} + dy^2\,,
\label{metric-dS}
\ee
%%%%%%%%%%%
and
%%%%%%%%%%%
%\be
%e^{A(y)}= \cosh (k y) - \frac{\kappa_5 \sigma}{\sqrt{6 |\Lambda_B|}}
%\,\sinh (k |y|)\,.
%\label{dS}
%\ee
\be
e^{A(y)}={\sinh[k(y_0-|y|)]\over \sinh(k y_0)}
\label{dS}
\ee
where here and in the following examples we shall consider $y_0>0$.

In the above, $\Lambda$ denotes the four-dimensional cosmological constant
given by
%%%%%%%%%%%%
\be
\Lambda = \frac{\kappa^2_5}{6}\,\Bigl(\frac{\kappa^2_5 \sigma^2}{6} -
|\Lambda_B|\Bigr)\,,
\ee
%%%%%%%%%%%%%%
where $\sigma$ is the (positive) tension of the brane located at $y=0$.
For $\kappa_5 \sigma > \sqrt{6 |\Lambda_B|}$, the effective cosmological
constant is positive, thus, leading to a de Sitter four-dimensional
spacetime on the brane. In this case, the warp factor vanishes at a
finite coordinate distance $y=y_0$ from the brane, with $y_0$ defined by
%%%%%%%%%%%%%%
\be
\tanh (k y_0) = \sqrt{\frac{6 |\Lambda_B|}{\kappa^2_5 \sigma^2}}\,.
\label{def-y0}
\ee
%%%%%%%%%%%%%%
The curvature invariant quantities $R$ and $R_{AB} R^{AB}$ are found to be
everywhere well-defined, however, Eq. (\ref{B}), for the square
of the Riemann tensor, reduces to
%%%%%%%%%%%%%
\be
R_{ABCD}\,R^{ABCD}= 40 k^4 + \frac{48 M^2 e^{-4 A(y)}}{r^6}\,,
\label{Riemann-dS}
\ee
%%%%%%%%%%%%%
which reveals the existence of a true singularity at $y=y_0$, and at the
location of the black-hole singularity, $r=0$.

Another solution, that falls in the same category, is the one that follows
if one combines the Schwarzschild-like metric tensor $\hat g_{\mu\nu}$
with the solution for the warp factor in the presence of a massless bulk
scalar field \cite{KOP1}. In this case, the five-dimensional metric
tensor has the form (\ref{metric-bs}) with the warp factor and the
scalar field assuming the form
%%%%%%%%%%%
\be
e^{4 A(y)}= a_0^4 \,\frac{\sinh [\omega\,(y_0-|y|)]}{\sinh (\omega\,y_0)}\,,
\qquad \quad
\phi'(y)=\frac{c\,a_0^4}{e^{4 A(y)}}\,.
\label{sc-field}
\ee
%%%%%%%%%%
In the above expressions,  $\omega^2=8 \kappa^2_5 |\Lambda_B|/3$,
while $a_0$ and $c$ are arbitrary constants. As in the case of the
4D de Sitter space, the warp factor vanishes at $y=y_0$, with $y_0$
being defined by
%%%%%%%%%%%%%%
\be
\sinh (\omega y_0) = \sqrt{\frac{2 |\Lambda_B|}{c^2}}\,.
\ee
%%%%%%%%%%%%%%
Turning to the Riemann curvature invariant, we find the result
%%%%%%%%%%%%%
\be
(R_{ABCD})^2= \frac{\omega^4}{256}\,\frac{\Bigl[207 -
44 \cosh[2 \omega\,(y_0-|y|)] + 5 \cosh[4 \omega\,(y_0-|y|)]\Bigr]}
{\sinh^4[\omega\,(y_0-|y|)]} + \frac{48 M^2 e^{-4 A(y)}}{r^6}\,,
\label{Riemann-sc}
\ee
%%%%%%%%%%%%%
which confirms the existence of a true singularity at
$y=y_0$ in addition to the black-hole singularity.

A similar procedure to the one discussed in the previous subsection needs
to be followed here for the resolution of the singularity problem.
The boundary singularity at $y=y_0$ can be shielded only by the
introduction of a second brane at $y=L<y_0$.
A {\it black tube} will again interpolate between the
two branes, whose four-dimensional line-elements will describe
dS-Schwarzschild black holes, thus giving a finite size to the
black-hole singularity. The monotonic profile of the warp factor
in the bulk demands a negative tension for the second brane and
unequal masses for the four-dimensional black-holes.

\bigskip

{\textbf{3. Non-vanishing, regular warp factor.}}

\medskip

The last category of solutions that we are going to study corresponds
to the class of five-dimensional line-elements (\ref{5-fact}) whose warp
factor is everywhere well-defined. The line-element (\ref{metric-dS}),
together with Eq. (\ref{dS}), describes a four-dimensional
Anti-de-Sitter spacetime on the brane when
$\kappa_5 \sigma < \sqrt{6 |\Lambda_B|}$. In that case, $\Lambda<0$
and the warp factor remains everywhere non-zero. The
point $y_0$ now denotes the location along the extra dimension where
the warp factor has a minimum. The square of the Riemann tensor is
given again by Eq. (\ref{Riemann-dS}), however, in this case, no
divergence appears at any point along the extra dimension and the
only singularity is the black-hole one.

Another solution, that exhibits a similar behaviour in the bulk, is the
one presented in Ref. \cite{KOP1}. The solution for the warp factor
in that case was given by
%%%%%%%%%%%
\be
e^{2 A(y)}= a_0^2 \,\cosh [\omega (|y|-y_0)]\,,
\label{T55}
\ee
%%%%%%%%%%
where $\omega^2=2 \kappa^2_5 |\Lambda_B|/3$,
and the brane is at $y=0$.
The metric tensor $\hat g_{\mu\nu}$ in the original model
was the Minkowski metric tensor, $\eta_{\mu\nu}$, however, the
implementation of the four-dimensional Schwarzschild line-element is
straightforward. The square of the Riemann tensor, in this case, is
found to be
%%%%%%%%%%%%%
\be
R_{ABCD}\,R^{ABCD}= \frac{\omega^4}{16}\,\frac{\Bigl[47 +
12 \cosh[2 \omega (|y|-y_0)] + 5 \cosh[4 \omega (|y|-y_0)]\Bigr]}
{\cosh^4[\omega (|y|-y_0)]} + \frac{48 M^2 e^{-4 A(y)}}{r^6}\,,
\label{Riemann-55}
\ee
%%%%%%%%%%%%%
which confirms the existence of only the black-hole singularity
at $r=0$.

Focusing on the second, singular term of Eqs. (\ref{Riemann-dS})
and (\ref{Riemann-55}), for the AdS ($\Lambda<0$) solution (\ref{dS})
and the solution (\ref{T55}), we notice that this term, together with
the black-hole singularity, vanishes exponentially fast with the
distance from the brane since, for $|y| \rightarrow \infty$, we obtain
%%%%%%%%%%%
\be
\frac{48 M^2 e^{-4 A(y)}}{r^6} \simeq \frac{const.}{r^6}\,
e^{-n\omega |y|}\,,
\ee
%%%%%%%%%%%%%
with $n=4$ for the AdS solution and $n=2$ for the solution (\ref{T55}),
while the first term reduces to a constant in both cases. Therefore,
for the above solutions, as one moves away from the brane, the
black-hole singularity `decays'.

The question that arises almost immediately is the behaviour of gravity for
those backgrounds in this case: the warp factor blows up at large distances
from the brane, a fact which casts doubts on the localization of gravity.
For the four-dimensional AdS spacetimes, it has been shown \cite{KR} that,
although no massless zero-mode exist in this case, a very light Kaluza-Klein
mode can play the role of the massless graviton and reproduce four-dimensional
gravity, provided that the effective cosmological constant, $\Lambda$,
is small.

Alternatively, the localization of gravity can be achieved by means of
the introduction of a second brane, as in the case of flat and dS branes.
There are actually a number of options concerning the location of the
second brane and thus the sign of its tension. If the second brane is
placed at $y=L<y_0$, that is on the {\it lhs} of the minimum, the
situation resembles the one for the flat and dS branes: its tension
must be negative and a {\it black tube} will interpolate between a
black hole with mass $M_1$ on the first brane and a black hole with
mass $M_2<M_1$ on the second; in addition, the black-hole singularity,
due to the decrease of the value of the warp factor, will get enhanced
as we move towards the second brane.

If $L=y_0$, the second brane necessarily has zero tension. In other words,
we may use the `turn' of the warp factor in order to compactify the
extra dimension without the introduction of a second brane \cite{kkop},
simply by identifying the two minima $y=\pm y_0$ on both sides
of the brane. In that case, a {\it black hole torus} would emerge
from the brane, traverse the five-dimensional bulk and return
to the brane. The black-hole singularity would get most enhanced
at the farthest point from the brane, that is at the minimum $y_0$,
while it would weaken as we moved towards the brane.

Finally, for $L>y_0$, that is for any brane placed on the {\it rhs} of
the minimum, the tension will always be positive, thus, circumventing
the necessity for the introduction of a, physically unrealistic,
negative-tension brane. For $L=2y_0$, the {\it black tube} connects
two four-dimensional black holes of {\it equal} masses, $M_1=M_2$.
Once we cross this critical point, $M_2$ will always be larger than  $M_1$.
Moreover, after this point, the black-hole singularity starts `decaying'
as the singular term, in the expression of the Riemann curvature
invariant, assumes smaller and smaller values compared to the one
in the neighbourhood of the first brane. Due to the exponential decaying
behaviour of this term, the black-hole singularity gets effectively
`localized' for even moderate distances of the second brane from
the location of the minimum $y_0$.

%%%%%%%%%%%%%%%%%%%%%%%%%%%%%%%%%%%%%%%%%%%%%%%%%%%%%%%%%%%%%%%%%%%%%%%%%%%%%%

\sect{The Higher-dimensional Case}

We now turn to the study of models with two or more extra dimensions.
As announced in the introduction, we are going to consider two particular
models, variants of previously derived six-dimensional brane-world solutions:
the Cohen-Kaplan string
solution \cite{CK} and the inhomogeneous solution of Ref. \cite{Kanti:2001vb}.
The higher-dimensional line-elements will be generalized in order to
include black-hole spacetime backgrounds on the brane and the analysis
presented in the previous section will be extended in the particular case.

\bigskip

{\textbf{{1. Cohen-Kaplan-Schwarzschild solution.}}

\medskip

Cohen and Kaplan \cite{CK} presented an interesting solution in
six spacetime dimensions describing a global 3-brane
modelled by a complex scalar field and zero bulk cosmological constant.
The solution can be considered as a higher-dimensional generalization of the
global string solutions. In the higher-dimensional
version of codimension 2, the core of the defect describes
a 3-brane instead of a 1-brane.
The existence of a bulk naked singularity, that is mild enough to admit
unitary boundary conditions, results in the compactification of the extra
spacetime leading to a finite radius for the transverse dimensions.

The specific form of the metric has $q$-dimensional Poincare invariance
in the directions parallel to the brane and is rotationally invariant in the
transverse space. Following the exact notation of Cohen and Kaplan, we write
as the metric of the model
\beq
ds^2=A^2(u)\,d\hat s^2_q +\gamma^2 B^2(u) (du^2+d\theta^2)\,,
\eeq
where $ d\hat s^2_q=\eta_\mn dx^{\mu} dx^{\nu}$, and $(u, \theta)$ stand
for the radial and angular coordinate, respectively, along the transverse
dimensions. The metric functions $A(u)$ and $B(u)$ turn out to be
\beq
A(u)=\left(u\over u_0 \right)^{1/q},
\q\q  B(u)=\left(u_0 \over u \right)^{(q-1)/2q}\;e^{(u_0^2-u^2)/2u_0}\,.
\eeq
This metric describes the solution outside the core
of the global brane defect that is located at $u\sim u_0$.
At $u=0$ there is a naked curvature singularity at a finite
proper distance from the core. Since $d\hat s^2_q$ is flat, we have
$\hat G_\mn=0$. But, as we have argued above,
any solution of the vacuum equations $\hat G_\mn=0$
will be a solution to the higher-dimensional Einstein equa\-tions
 with the same functions $A(u)$ and $B(u)$.
A natural solution to consider along the longitudinal directions
is that of a spherical Schwarzschild black hole
%%%
\be
d\hat{s}^2_{Schw}=-h(r)\,dt^2+{dr^2\over h(r)}+r^2 d\Omega_2^2\,,
\ee
with $h(r)=1-2M/r$. This solution has
\be \hat{R}_{(4)}=0, \q\q (\hat{R}_{\mn \alpha
\beta})^2={48\,M^2\over r^6}\,.\ee
In the higher-dimensional spacetime, the expression for the square of
the Riemann tensor comes out to be
%%%%%
\be
R_{ABCD}\,R^{ABCD}= \frac{3 B^4(u)\,(32 u^4 -16 u^2 u_0 + 5 u_0^2)}
{16 \gamma^4\,u_0^5\,u} + \frac{48\,M^2}{A^4(u)\,r^6}\,.
\ee
%%%%%%%%%
As it is obvious from the second term in the above expression, the black
hole singularity, located at $r=0$, is indeed extended over the two
dimensional transverse space, however this is protected by a horizon.
The same term reveals the appearance of naked singularities at the points,
along the $u$-dimension, where the warp factor vanishes.
However, since the only zero of the warp factor is at $u=0$, this is
not going to be a new singularity but will coincide with the standard
singularity of the original Cohen-Kaplan solution (as the first term
shows, this singularity was still present even in the flat 4D spacetime
limit $M \rightarrow 0$). The situation is similar to the 5-dimensional
brane-world models with a
singularity in the bulk: due to the cylindrical symmetry of the
transverse spacetime, the warp factor is governed by only one
extra coordinate, the radial one. Any singularities appearing in
the case of flat brane line-elements remain unchanged when black-hole
line-elements are introduced, and the introduction of a 4-brane at $u=L>0$
provides the only way of shielding the naked singularity out of the
causal spacetime.

In the remainder of this subsection we will look at this naked singularity
in the case of a single-brane model and establish whether its nature
is changed by the introduction of the black hole on the brane or it can
still be considered harmless as in the original work by Cohen and Kaplan.
By following their analysis, we consider a more appropriate radial
coordinate and move to the {\em conformal} gauge. Defining $z$ by
\be dz=\gamma\,\frac{B(u)}{A(u)}\,du\,, \ee
we bring the metric to the form
\beq
ds^2= A^2(z)\,(ds_{Schw}^2+dz^2) +\gamma^2 B^2(z)\,d\th^2\,.
\eeq
Near the naked singularity, at $z=0$, the metric functions behave as
$A(z) \rightarrow z^{2/3}$ and $B(z)\rightarrow z^{-1}$.
In order to have a {\textit{harmless}} naked singularity, we require
the vanishing of the flow (\ref{flow}).

In \cite{CK}, it was shown that unitary boundary conditions
can be imposed for gravitational and scalar modes so that no
conserved quantity leaks through the naked singularity. In the
present case of a geometry with a black hole,
the solutions are still separable, so we can introduce a scalar field
\be
\Phi(x, z, \th)= \Phi_x(x)\,\Phi_z(z)\,e^{in\theta}\,.
\ee
%%%
Rewriting the $z$-dependent part of the scalar field as
%%%
\be
\Phi_z(z)=\frac{\vf(z) }{\psi(z)}\,, \quad \quad
\psi(z)\equiv (\gamma A^3B)^{1/2}\,,
\ee
we find -- not surprisingly-- that it satisfies the same equation
as in the Cohen-Kaplan case:
\beq
\left[-{d^2\over dz^2}+{\psi''\over \psi}+n^2 {A^2\over B^2} \right]
\vf(z)=m^2 \vf(z)\,.
\eeq
The following analysis and results are identical to those of Ref. \cite{CK}:
from the behaviour of $A$ and $B$, as $z\rightarrow 0$, we can see that,
near the singularity, the $n^2$ and $m^2$ terms are sub-dominant, so that
all the modes behave as the zero modes in this region. These modes can
be easily found by solving the above equation and their behaviour, as
$z \rightarrow 0$, is $\Phi_z^{(1)}(z)\sim 1$ and
$\Phi_z^{(2)}(z)\sim \log z$. The condition of vanishing flow now is
%%%%%
\be
\mathop{\lim}\limits_{z\to 0}\;\;\sqrt{-g} g^{zz}\Phi_z(z)\,\Phi_z'(z)
=\mathop{\lim}\limits_{z\to 0}\;\; z\,\Phi_z(z)\,\Phi_z'(z) =0\,,
\ee
%%%%
and is clearly satisfied only by $\Phi_z^{(1)}(z)$ thus excluding the modes
of the second kind from the set of physically acceptable solutions. We thus
see that there is no dramatic change in the CK solution when we introduce
the intersecting black disc in the bulk. The only difference is that the
$x^\mu$-dependent part of the scalar field must now satisfy Eq. (\ref{eq-4D})
in a four-dimensional Schwarzschild background instead of a Minkowski one
and a regular behaviour should be demanded near the black hole horizon.
We therefore conclude that, in the case of the Cohen-Kaplan-Schwarzschild
background, both massless and massive scalar modes, with the same trivial
behaviour at the vicinity of the naked singularity, are allowed to propagate
in the bulk without loss of unitarity.

%%%%%%%%%%%%%%%%%%%%%%%%%%%%%%%%%%%%%%%%%%%%%%%%%%%%%%%%%%%%%

\bigskip

{\textbf{2. Codimension Two Black Branes.}}

\medskip

In what follows, we shall consider a different class of codimension two
metrics, based on the six dimensional metrics presented in Ref.
\cite{Kanti:2001vb}.
Unlike the previous example the warp factor depends on both transverse
coordinates $\varphi$ and $\theta$, and therefore the extra spacetime
does not exhibit any cylindrical symmetry. We may write this metric as
\beq
ds^2=g^2(\vf)\cosh^2 [\lambda\,(\th-\th_m)]\,d\hat{s}_q^2
+\left(d\th^2+{1\over \lambda ^2}\cosh^2[\lambda\,(\th-\th_m)]
\,d\vf^2\right),
\label{kmo}
\eeq
where $\lambda$ is defined as
%%%%%%%%%%
\be
{\lambda^2}=-{2 \Lambda_B \over q(q+1)}
\ee
%%%%%%%%%%%
and it is thus associated with the radius of curvature of the $AdS_{q+2}$
spacetime. On the other hand, $\th_m$ gives the location of the minimum
of the warp factor along the $\th$-coordinate.
The generalization with respect to  \cite{Kanti:2001vb} is that we take
$q$ arbitrary while in \cite{Kanti:2001vb} it was set equal to four.
The total dimensionality of spacetime is thus $d=q+2$. The form of
the function $g(\vf)$ depends on the effective cosmological constant
required by the metric $d\hat{s}_q^2$. Thus,
%%%%%%%%%
\beqa
g(\vf)&=&{\sinh\left\{\vf_0-|\vf|\right\} \over \sinh(\vf_0)} \q\q\q
(\Lambda_q >0)\\
&=&e^{- |\vf|}  \q\q\q\q\q\q\q\q\,\,  (\Lambda_q=0) \\
&=&{\cosh\left\{|\vf|-\vf_0 \right\}  \over \cosh(\vf_0)}
\q\q\q (\Lambda_q<0)
\eeqa
Normalization is chosen so that $g(0)=1$ and we choose
$\vf_0>0$. We note that the metrics (\ref{kmo})
with $g(\vf)$ given by the above functions and $d\hat s_q^2$ by the
associated  maximally symmetric spacetimes
correspond to three different
foliations  of $AdS_{q+2}$. The curvature radius of the $q$-dimensional metric
is given by
\beq
\hat H^2 \equiv \frac{2 |\Lambda_q|}{(q-1)(q-2)}=
\biggl({\lambda^2\over \sinh^2(\vf_0)},
0, {\lambda^2\over \cosh^2(\vf_0)}\biggr)\,,
\eeq
for $\Lambda_q>0$, $\Lambda_q=0$, and $\Lambda_q<0$, respectively.
We also note that, for $\Lambda_q >0$ and $\Lambda_q =0$, the warp
factor vanishes at some point away from the brane. This signals the
presence of a horizon at a finite proper distance ($\Lambda_q >0$)
or at an infinite proper distance ($\Lambda_q =0$) from the brane .

In Ref. \cite{Kanti:2001vb}, $d\hat{s}_q^2$ corresponded to maximally
symmetric line elements and a 4-brane was introduced at $\vf=0$ by the
usual procedure of imposing symmetric configurations under the
transformation $\vf \leftrightarrow -\vf$.
A positive tension brane is obtained for warp factors that decrease in
the vicinity of the brane.
Instead of the maximally symmetric choice for the longitudinal part of the
metric, we may now proceed to consider black hole solutions along the brane.
To construct the black brane version of the above solutions,
we can consider for the $q$-dimensional part of the metric the
following ansatz
\beq
d\hat{s}^2_q=-h(r)\,dt^2+{dr^2\over h(r)}+r^2 d\Omega_{(q-2)}^2\,,
\eeq
with
\beq  h(r)=1-\left({2M\over r}\right)^{(q-3)}+\epsilon \hat H^2 r^2\,.\eeq
In this case, we have
\beq \hat{R}_{(q)}=-\epsilon q(q-1) \hat H^2, \q\q
 (\hat{R}_{\mn \alpha
\beta})^2=\frac{4(q-1)(q-2)^2(q-3)M^{2(q-3)}}{r^{2(q-1)}}+2q(q-1)\hat H^4\,.
\eeq
These metrics represent pure Schwarzschild ($\epsilon=0$),
Schwarzschild-de Sitter ($\epsilon=-1$) and Schwarzschild-anti de
Sitter ($\epsilon=1$) black holes in a $q$-dimensional spacetime.

Not surprisingly the introduction of the black brane gives rise
to an extended naked singularity. We can see this
from the expression of
the square of the Riemann tensor for these
$(q+2)$ dimensional metrics
\beq(R_{ABCD})^2=2 (q+1)(q+2)\lambda^4+
\frac{4(q-1)(q-2)^2(q-3)M^{2(q-3)}}
{\cosh^4[\lambda(\th-\th_m)]\,g^4(\vf)\,r^{2(q-1)}}\,. \eeq
Here, we can see explicitly how the horizons become singular. Any
zero of  $g(\vf)$ signals a naked singularity. For the particular
solutions under consideration $g(\vf)$ goes to zero for
$\Lambda_q >0$ as $\vf \rightarrow \vf_0$ and for $\Lambda_q =0$
as $|\vf| \rightarrow \infty$.
For $\Lambda_q <0$, there is no new singularity since
the warp factor is always different from zero.
If we keep the $\theta$-coordinate fixed, then, the profile of the
warp factor along the $\varphi$-coordinate is similar to the one
in the case of the five-dimensional spacetimes considered in
Section 3: for $\Lambda_q$ zero or positive, the higher-dimensional
spacetime is characterized by two bulk singularities, one covered
by a horizon, the other naked; in order to shield the second singularity
and give a finite size to the first one, we need to introduce a second
brane at $\varphi=\varphi_L$, before the zero of the warp factor takes
place. For $\Lambda_q$ negative, the warp factor along the
$\varphi$-coordinate has a cosh-like behaviour which causes the
suppression of the black hole singularity for distances
$\varphi>\varphi_0$, i.e. for distances moderately larger than
the location of the minimum of the warp factor. In this last case,
therefore, no brane needs to be introduced as the black hole singularity
rapidly decays away from the brane, while an effective
compactification may take place through the mechanism of Ref. \cite{KR}.

If we now allow for the $\theta$-coordinate to vary as well, then an
additional `localizing' effect appears. Due to the absence of any
cylindrical symmetry, as opposed to the model studied in the previous
subsection, the warp factor depends on both transverse coordinates
and thus an extra cosh-like function appears in the denominator of
the singular term in the expression of $(R_{ABCD})^2$. The $\th$-profile
of the warp factor remains the same for every value of $\Lambda_q$
and leads to the suppression of this singular term at distances
$\theta > \theta_m$, independently of the form of the $\vf$-profile.
%Thus, even in the case of single-brane configurations, both the
%black hole singularity and the naked singularity (for $\Lambda_q \geq 0$)
%disappear if one moves along the 4-brane and towards large values
%of $\theta$.
Due to the cosh-like profile along this dimension,
an effective compactification may also take place as in Ref. \cite{KR}.

We next consider the flow into the naked singularity for scalar field modes
and check that no conserved quantity leaks through it.
The form of the metric leads again to separable equations. So
we can write the scalar field mode as
$\Phi(x,\theta,\vf)=\Phi_x(x)\,\Phi_\th(\th)\,\Phi_\vf(\vf)$.
The flow vanishing condition (\ref{flow}) now leads to
\beq
\mathop{\lim}\limits_{\vf \to \vf_0} \;\;
g^q(\vf)\,\Phi_\vf(\vf)\,\Phi_\vf'(\vf)
=0\,. \label{cond}
\eeq
Thus, we need only the explicit behavior along
the $\vf$ direction close to the singularity.
The separated equations for the scalar modes along the extra spacetime are
\beq g^{-(q-2)}\p_{\vf}(g^q \;\p_{\vf} \Phi_\vf) +\Phi_\vf(m^2-k g^2)=0\,,\eeq
\beq f^{-(q-1)}\p_\th(f^{(q+1)} \;\p_\th \Phi_\th)+k\Phi_\th =0\,,\eeq
where $m^2$ is the mass of the modes from the 4D perspective,
$k$ is a separation constant and $f=\cosh[\lambda\,(\th-\th_m)]$.
Unlike the Cohen-Kaplan example, near the singularity, when $g \rightarrow 0$,
the massive term in the first equation is not sub-dominant, although the
$k-$term is.

It is convenient to change the independent
variable so that $dz=d\vf /\lambda g(\vf)$ and
rescale
 \beq
\Phi_z(z)=\frac{\chi(z)}{\Psi(z)}\,\,,\,\,\,\,\,\,\,
\Psi(z)=g(z)^{(q-1)/2}\,,
\eeq
so that the equation takes the form of a quantum mechanics
problem\footnote{
Note that the normalization for the modes comes from the condition
$$\int\sqrt{g}\,g^{AB}\,\p_A\Phi\,\p_B\Phi
\supset
\int\sqrt{g}\,(g(\vf)\,\cosh(\lambda \th))^{-2}\,
\hat g^{\mn}\,\p_{\mu}\phi\,\p_{\nu} \phi
\sim \int  \chi(z)^2\;dz<\infty\,.$$}
%%%%%%%%%%%%%%%
\beq \left[ -{d^2\over dz^2}\;+V\right]\chi
=m^2\chi, \q\q\q  V\equiv {\Psi''\over \Psi} +k\,g^2\,,
\eeq
where primes now stand for $\p_z$.
In the limit $g\rightarrow 0$, we can safely drop the
$k-$term.

We first address the $\Lambda_q >0$ case where
$g\propto \sinh(\vf_0-|\vf|)=1/\sinh (H (z_0+|z|))$ so
the limit $\vf \rightarrow \vf_0$ corresponds to
$z\rightarrow \infty$.  Due to the reflection symmetry across the wall,
we will consider only the singularity at one side of the wall.
In the limit $z \rightarrow \infty$, we have
\beq
-{d^2\over dz^2}\;\chi=\left[m^2-{(q-1)^2\over 4}\;\lambda^2 \right]\chi\,.
\eeq
An approximate solution near the singularity is given in terms of
rising and falling exponentials
\beq
\chi_{\pm}(z)=e^{\pm \Delta z}\,,
\eeq
with $\Delta\equiv \sqrt{{(q-1)^2 \lambda^2/4}-m^2}$.
The flow into the singularity (\ref{cond}) for these two modes is
proportional to
\beq
\mathop{\lim}\limits_{z \to \infty} g^{(q-1)}(z)\,\Phi_z(z)\,\Phi_z'(z)=
\mathop{\lim}\limits_{z \to \infty}\left[
 e^{2\Delta z}, e^{-2\Delta z}\right]
\eeq
for $\chi_{+}$ and $\chi_{-}$, respectively.
Note that there is a critical value
$m_c^2\equiv (q-1)^2 \lambda^2/4$. For $0\leq m^2< m_c^2$, $\Delta$ is real and
it is clear that only the mode given by $\chi_{-}$ is physical, leading to
a vanishing flow through the singularity.
For $m^2> m_c^2$, the modes are proportional to plane waves and the
limit corresponds to a wild oscillation, leading again to a vanishing flow.

One can wonder about the existence of modes with $m^2< m_c^2$. After all
$m_c^2$ is the constant value of the
potential as $z \rightarrow \infty$ and we know
that, at best, we will have a single normalizable mode when
$m_c^2$ is not the absolute minimum of the potential.
It is natural thus to look for those modes and see whether
they coincide with the vanishing flow modes.
For this we have to consider
the full expression of the potential for $\chi$
\beq
V={(q-1)^2 \lambda^2\over 4}+{(q^2-1)\lambda^2
+4 k \sinh^2(\lambda z_0) \over 4\sinh^2(\lambda (z_0+|z|))}
-(q-1)\lambda \coth (\lambda z_0) \delta(z)
\eeq

By looking at the equation along $\th$, one can see
that $k>0$, so this is basically the known volcano
potential for a de Sitter brane in $AdS_5$. We recover
that expression for $q=4$, $k=0$. The delta function
due to the presence of the wall supports the zero mode
and then we have the continuum of KK modes with
the usual gap $m^2>m_c^2$. Since the KK modes
lie at $m^2>m_c^2$, they clearly satisfy the
vanishing flow condition. We only have to check that
the zero mode with the vanishing flow is normalizable and
thus that it corresponds to the localized graviton.
The zero mode with vanishing flow
is approximated by $\chi(z)\sim e^{-\lambda|z|(q-1)/2}$
as $|z|\rightarrow \infty$, so the integral for $\chi^2(z)$
will converge. We have thus shown that none of the normalizable
modes leaks through the singularity.

In the $\Lambda_q =0$ case,
$g(\vf)= e^{-|\vf|}=1/(1+ \lambda |z|)$. In the new
parametrization  the singularity is approached when
$z \rightarrow \infty$, and the equation for
the scalar modes can be approximated by
\beq
\left[-{d^2\over dz^2}+{(q^2-1)\lambda^2+4k \over 4 \lambda^2  }
\;{1\over z^2}
\right]\;\chi= m^2\chi\,.
\label{Lq=0}
\eeq
For massless modes, close to the singularity we have
\beq
\chi_{\pm}(z)\sim
(\lambda z)^{(1\pm \sqrt{q^2+4k/\lambda^2})/2}\,,
\eeq
and the flow is proportional to
\beq
\mathop{\lim}\limits_{z \to \infty} g^{(q-1)}(z)\,\Phi_z(z)\,\Phi_z'(z)
\sim \mathop{\lim}\limits_{z \to \infty}
\left[ (\lambda z)^{\sqrt{q^2+4k/\lambda^2}}, (\lambda z)^{-\sqrt{q^2+4k/\lambda^2}}\right]
\eeq
for $\chi_{+}$ and $\chi_{-}$, respectively.
Clearly, $\chi_{-}$
%One of the profiles
corresponds to the acceptable massless mode
with vanishing flow through the singularity.
This mode is normalizable since
\beq
\int_0^{\infty} \chi_{-}^2(z) \;dz={1\over \sqrt{\lambda^2 q^2+4k}}\,.
\eeq
Finally, for massive modes, we can drop the term in $1/z^2$ in Eq. (\ref{Lq=0})
and we find again plane waves
\beq \chi(z)=
C_1 e^{imz}+C_2 e^{-imz}\,,  \eeq
which lead to vanishing flow through the singularity
and can be normalized in the usual way.

For completeness, we again give the full potential which now has the form
\beq
V={(q^2-1)\lambda^2
+4k  \over 4(1+\lambda |z|)^2}
-(q-1)\lambda  \delta(z)\,.
\eeq
Again, for $q=4$ and $k=0$, this reduces to the volcano potential for a
flat brane in $AdS_5$. There is thus a normalizable zero mode supported
by the delta function and the usual continuum of KK modes.
As we have shown above, all of these modes have vanishing flow.
We have therefore shown that, in the present example, the vanishing flow
condition doesn't lead to additional constraints, since it is satisfied
by all normalizable modes. This suggests that the naked singularity is
practically harmless from this point of view, and unitarity constraints
are still satisfied by scalar modes propagating in the bulk.

{\section{Conclusions}}

The objective of this paper has been the investigation of the singular
properties of higher-dimensional brane-world spacetimes when a
spherically-symmetric Schwarzschild-like black-hole background is
introduced on the brane. The appearance of old and new singularities,
covered by a horizon or unshielded, has been studied by computing the
corresponding curvature invariant quantities of those spacetimes.
Ways of shielding or suppressing these singularities have been
discussed and the possible flow of conserved currents into the
naked singularities has also been studied.

In the first part of this paper, we considered various examples of factorizable
five-dimensional geometries all of which were associated with Schwarzschild
black string solutions. Brane-world models with an exponential or sinh-like
warp factor were additionally plagued by the presence of a naked singularity
in the bulk. In these cases, the introduction of a second brane provides
the only way of shielding the bulk singularity and generating a finite-size
black-tube construction.  On the other hand, brane-world solutions with a
cosh-like warp factor are free from the bulk singularity, and
we can construct brane
configurations  with either a compact or non-compact extra dimension.
In both cases, the black string becomes effectively
localized near the brane: as the warp factor increases away from the brane,
the corresponding singular term, in the expression of the square of the
Riemann tensor, gets suppressed at moderate distances
from the brane.

We further moved, in the second part of the paper, to consider codimension-2 models
of the same factorizable type. In particular, we considered a well-known
``cylindrically symmetric" model, the Cohen-Kaplan model, and the ``asymmetric"
six-dimensional model of Ref. \cite{Kanti:2001vb}, both appropriately modified
in order to include a black hole line-element on the brane. These models admit
infinitely-extended black brane solutions with similar features to the
Schwarzschild black-string solution. The variant of the CK model is also
characterized by the same naked singularity that was present in the original
version - if we want this singularity shielded, we have no other choice but to
introduce a second brane.
%, which gives a finite size also to the black brane.
The second model accepts a variety of solutions depending on the value of the
effective cosmological constant on the brane: flat and positively-curved branes
lead to solutions with a naked bulk singularity, while negatively-curved branes
are bulk-singularity-free. In the first two cases, the naked singularity can be
avoided and the black brane can be given a finite size only through the
introduction of a second brane. In the latter case, however, the cosh-like
form of the warp factor may lead once again to the effective localization of
the black brane construction in a similar way to the five-dimensional case.
This ``asymmetric" model is characterized by an additional localizing effect
that comes from the dependence of the warp factor on both extra dimensions:
another cosh-like function, present for all values of the effective cosmological
constant, suppresses the singular terms in the square of the Riemann tensor
as one moves parallel to the brane, and thus leads to the ``decaying" of both
the black-hole and the bulk singularity.

Single-brane configurations may be constructed if the bulk singularity is left
unshielded. In this case, it is the singularity itself that terminates the
extra dimension and gives a finite size to the black string or black brane.
As suggested in the original analysis by Cohen and Kaplan, this type of
singularities could be considered harmless if no flow of conserved quantities
takes place through them. By considering an auxiliary bulk scalar field to
realize a conserved current associated with the spacetime symmetries, we
have studied a variety of models, five-, six- or higher-dimensional, and we
have showed that it is always possible to choose appropriate particle modes
in order to satisfy the constraint of vanishing flow. Attempts to construct
brane-world models in five dimensions, where the naked singularity was used
in the above way, resulted into solutions \cite{self} that had either a severe,
hidden fine-tuning \cite{fine}\cite{KOP1} or a badly-defined effective
four-dimensional theory \cite{eff} (see also \cite{cline}). A particular
class of five-dimensional
models that was studied here, the one with a bulk scalar field, falls in this
category and therefore the shielding of the naked singularity is imperative
- the introduction of the black-hole line-element on the brane, in this
case, did not affect the naked singularity, which pre-existed. However, we
have also considered alternative models, both five- and higher-dimensional ones,
where the bulk singularity was induced merely by the introduction of the black
hole on the brane while the bulk itself was everywhere well-defined. In these
cases, it remains to be seen whether the ill-defined behaviour, encountered
in some five-dimensional models, also persists here or whether it can be
ameliorated in the framework of more complex brane-world models.

\bigskip

{\bf Acknowledgements.}
I.O. and K.T. acknowledge the financial support of the EU RTN contract
No. HPRN-CT-2000-00152.

%%%%%%%%%%%%%%%%%%%%%%%%%%%%%%%


\begin{thebibliography}{99}


\bibitem{large1}
I.~Antoniadis,
%``A Possible New Dimension At A Few Tev,''
Phys.\ Lett.\ B {\bf 246}, 377 (1990) ;
%%CITATION = PHLTA,B246,377;%%
%%
N.~Arkani-Hamed, S.~Dimopoulos and G.~R.~Dvali,
%``The hierarchy problem and new dimensions at a millimeter,''
Phys.\ Lett.\ B {\bf 429}, 263 (1998)
[hep-ph/9803315];
%%CITATION = HEP-PH 9803315;%%
%%
I.~Antoniadis, N.~Arkani-Hamed, S.~Dimopoulos and G.~R.~Dvali,
%``New dimensions at a millimeter to a Fermi and superstrings at a TeV,''
Phys.\ Lett.\ B {\bf 436}, 257 (1998)
[hep-ph/9804398].
%%CITATION = HEP-PH 9804398;%%


\bibitem{RS1}
L.~Randall and R.~Sundrum,
%``A large mass hierarchy from a small extra dimension,''
Phys.\ Rev.\ Lett.\  {\bf 83}, 3370 (1999)
[hep-ph/9905221].
%%CITATION = HEP-PH 9905221;%%

\bibitem{RS2}
L.~Randall and R.~Sundrum,
%``An alternative to compactification,''
Phys.\ Rev.\ Lett.\  {\bf 83}, 4690 (1999)
[hep-th/9906064].
%%CITATION = HEP-TH 9906064;%%

\bibitem{graviton}
J.~Lykken and L.~J.~Randall,
JHEP {\bf 0006} (2000) 014 [hep-th/9908076];
%%CITATION = HEP-TH 9908076;%%
K.~Skenderis and P.~K.~Townsend,
Phys.\ Lett.\ B {\bf 468} (1999) 46 [hep-th/9909070];
%%CITATION = HEP-TH 9909070;%%
K.~Behrndt and M.~Cvetic, Phys.\ Lett.\ B {\bf 475} (2000) 253
[hep-th/9909058];
%%CITATION = HEP-TH 9909058;%%
A.~Chamblin and G.~W.~Gibbons,
Phys.\ Rev.\ Lett.\  {\bf 84} (2000) 1090 [hep-th/9909130];
%%CITATION = HEP-TH 9909130;%%
O.~DeWolfe, D.~Z.~Freedman, S.~S.~Gubser and A.~Karch,
Phys.\ Rev.\ D {\bf 62} (2000) 046008 [hep-th/9909134];
%%CITATION = HEP-TH 9909134;%%
C.~Csaki, J.~Erlich, T.~J.~Hollowood and Y.~Shirman,
Nucl.\ Phys.\ B {\bf 581} (2000) 309 [hep-th/0001033];
%%CITATION = HEP-TH 0001033;%%
S.~B.~Giddings, E.~Katz and L.~J.~Randall, JHEP {\bf 0003} (2000) 023
[hep-th/0002091];
%%CITATION = HEP-TH 0002091;%%
H.~Kudoh and T.~Tanaka,
Phys.\ Rev.\ D {\bf 64} (2001) 084022 [hep-th/0104049];
%%CITATION = HEP-TH 0104049;%%
A. Kehagias and K. Tamvakis, hep-th/0205009.

\bibitem{CHR}
A.~Chamblin, S.~W.~Hawking and H.~S.~Reall,
Phys.\ Rev.\ D {\bf 61} (2000) 065007 [hep-th/9909205].
%%CITATION = HEP-TH 9909205;%%

\bibitem{Giannakis}
I.~Giannakis and H.~c.~Ren,
Phys.\ Rev.\ D {\bf 63} (2001) 024001 [hep-th/0007053].
%%CITATION = HEP-TH 0007053;%%

\bibitem{GL}
R.~Gregory and R.~Laflamme,
Phys.\ Rev.\ Lett.\  {\bf 70} (1993) 2837 [hep-th/9301052].
%%CITATION = HEP-TH 9301052;%%

\bibitem{bh-brane}
R.~Emparan, G.~T.~Horowitz and R.~C.~Myers,
JHEP {\bf 0001} (2000) 007; [hep-th/9911043]
%%CITATION = HEP-TH 9911043;%%
J.~Garriga and M.~Sasaki, Phys.\ Rev.\ D {\bf 62} (2000) 043523
[hep-th/9912118];
%%CITATION = HEP-TH 9912118;%%
A.~Chamblin, C.~Csaki, J.~Erlich and T.~J.~Hollowood,
Phys.\ Rev.\ D {\bf 62} (2000) 044012 [hep-th/0002076];
%%CITATION = HEP-TH 0002076;%%
N.~Dadhich, R.~Maartens, P.~Papadopoulos and V.~Rezania,
Phys.\ Lett.\ B {\bf 487} (2000) 1 [hep-th/0003061];
%%CITATION = HEP-TH 0003061;%%
S.~Nojiri, O.~Obregon, S.~D.~Odintsov and S.~Ogushi,
Phys.\ Rev.\ D {\bf 62} (2000) 064017 [hep-th/0003148];
%%CITATION = HEP-TH 0003148;%%
A.~Chamblin, H.~S.~Reall, H.~a.~Shinkai and T.~Shiromizu,
Phys.\ Rev.\ D {\bf 63} (2001) 064015 [hep-th/0008177];
%%CITATION = HEP-TH 0008177;%%
I.~Giannakis and H.~c.~Ren,
Phys.\ Rev.\ D {\bf 63} (2001) 125017 [hep-th/0010183];
%%CITATION = HEP-TH 0010183;%%
Phys.\ Rev.\ D {\bf 64} (2001) 065015 [hep-th/0103265].
%%CITATION = HEP-TH 0103265;%%
M.~Bruni, C.~Germani and R.~Maartens,
%``Gravitational collapse on the brane,''
Phys.\ Rev.\ Lett.\  {\bf 87}, 231302 (2001) [gr-qc/0108013];
%%CITATION = GR-QC 0108013;%%
M.~Rogatko, Phys.\ Rev.\ D {\bf 64} (2001) 064014 [hep-th/0110018];
%%CITATION = HEP-TH 0110018;%%
I.~Giannakis and H.~c.~Ren,
%``Linearized analysis of the Dvali-Gabadadze-Porrati brane model,''
Phys.\ Lett.\ B {\bf 528}, 133 (2002) [hep-th/0111127];
%%CITATION = HEP-TH 0111127;%%
R.~Casadio, A.~Fabbri and L.~Mazzacurati,
%``New black holes in the brane-world?,''
Phys.\ Rev.\ D {\bf 65}, 084040 (2002) [gr-qc/0111072];
%%CITATION = GR-QC 0111072;%%
S.~I.~Vacaru and D.~Singleton,
%``Warped solitonic deformations and propagation of black holes in 5D
%vacuum gravity,''
Class.\ Quant.\ Grav.\  {\bf 19}, 3583 (2002) [hep-th/0112112];
%%CITATION = HEP-TH 0112112;%%
%%
R.~Emparan, A.~Fabbri and N.~Kaloper,
%``Quantum black holes as holograms in AdS braneworlds,''
hep-th/0206155.
%%CITATION = HEP-TH 0206155;%%


\bibitem{branes-bh}
C.~Gomez, B.~Janssen and P.~J.~Silva,
JHEP {\bf 0004} (2000) 027 [hep-th/0003002];
%%CITATION = HEP-TH 0003002;%%
A.~Kamenshchik, U.~Moschella and V.~Pasquier,
Phys.\ Lett.\ B {\bf 487} (2000) 7 [gr-qc/0005011];
%%CITATION = GR-QC 0005011;%%
D.~Youm, Phys.\ Lett.\ B {\bf 515} (2001) 170 [hep-th/0105093];
%%CITATION = HEP-TH 0105093;%%
Mod.\ Phys.\ Lett.\ A {\bf 16} (2001) 1703 [hep-th/0107174];
%%CITATION = HEP-TH 0107174;%%
C.~Grojean, F.~Quevedo, G.~Tasinato and I.~Zavala C.,
JHEP {\bf 0108} (2001) 005 [hep-th/0106120];
%%CITATION = HEP-TH 0106120;%%
S.~Nojiri, S.~D.~Odintsov and S.~Ogushi,
%``Holographic entropy and brane FRW dynamics from AdS black hole in
% d5 higher derivative gravity,''
Int.\ J.\ Mod.\ Phys.\ A {\bf 16}, 5085 (2001) [hep-th/0105117];
%%CITATION = HEP-TH 0105117;%%
R.~G.~Cai and Y.~Z.~Zhang,
%``Holography and brane cosmology in domain wall backgrounds,''
Phys.\ Rev.\ D {\bf 64}, 104015 (2001) [hep-th/0105214];
%%CITATION = HEP-TH 0105214;%%
D.~Birmingham and M.~Rinaldi,
%``Brane world in a topological black hole bulk,''
Mod.\ Phys.\ Lett.\ A {\bf 16}, 1887 (2001) [hep-th/0106237].
%%CITATION = HEP-TH 0106237;%%



\bibitem{super}
H.~Lu and C.~N.~Pope,
Nucl.\ Phys.\ B {\bf 598} (2001) 492 [hep-th/0008050];
%%CITATION = HEP-TH 0008050;%%
C.~V.~Johnson and R.~C.~Myers,
Phys.\ Rev.\ D {\bf 64} (2001) 106002 [hep-th/0105159];
%%CITATION = HEP-TH 0105159;%%
S.~S.~Gubser, A.~A.~Tseytlin and M.~S.~Volkov,
JHEP {\bf 0109} (2001) 017 [ hep-th/0108205].
%%CITATION = HEP-TH 0108205;%%

\bibitem{stability}
H.~S.~Reall,
Phys.\ Rev.\ D {\bf 64} (2001) 044005 [hep-th/0104071];
%%CITATION = HEP-TH 0104071;%%
G.~T.~Horowitz and K.~Maeda,
Phys.\ Rev.\ Lett.\  {\bf 87} (2001) 131301 [hep-th/0105111];
%%CITATION = HEP-TH 0105111;%%
J.~P.~Gregory and S.~F.~Ross,
%``Stability and the negative mode for Schwarzschild in a finite cavity,''
Phys.\ Rev.\ D {\bf 64}, 124006 (2001) [hep-th/0106220];
%%CITATION = HEP-TH 0106220;%%


\bibitem{higher}
M.~R.~Mbonye, Phys.\ Rev.\ D {\bf 60} (1999) 124007 [gr-qc/9908054];
%%CITATION = GR-QC 9908054;%%
S.~W.~Hawking and H.~S.~Reall,
Phys.\ Rev.\ D {\bf 61} (2000) 024014 [hep-th/9908109];
%%CITATION = HEP-TH 9908109;%%
P.~Kraus, JHEP {\bf 9912} (1999) 011 [hep-th/9910149];
%%CITATION = HEP-TH 9910149;%%
A.~M.~Awad and C.~V.~Johnson,
Phys.\ Rev.\ D {\bf 63} (2001) 124023 [hep-th/0008211];
%%CITATION = HEP-TH 0008211;%%
C.~Cadeau and E.~Woolgar,
Class.\ Quant.\ Grav.\  {\bf 18} (2001) 527 [gr-qc/0011029];
%%CITATION = GR-QC 0011029;%%
R.~Emparan and H.~S.~Reall,
%``A rotating black ring in five dimensions,''
Phys.\ Rev.\ Lett.\  {\bf 88}, 101101 (2002) [hep-th/0110260];
%%CITATION = HEP-TH 0110260;%%
U.~H.~Danielsson,
%``A black hole hologram in de Sitter space,''
JHEP {\bf 0203}, 020 (2002) [hep-th/0110265].
%%CITATION = HEP-TH 0110265;%%

\bibitem{large}
P.~C.~Argyres, S.~Dimopoulos and J.~March-Russell,
Phys.\ Lett.\ B {\bf 441} (1998) 96 [hep-th/9808138];
%%CITATION = HEP-TH 9808138;%%
R.~Emparan, G.~T.~Horowitz and R.~C.~Myers,
Phys.\ Rev.\ Lett.\  {\bf 85} (2000) 499 [hep-th/0003118];
%%CITATION = HEP-TH 0003118;%%
R.~Casadio and B.~Harms, Phys.\ Lett.\ B {\bf 487} (2000) 209
[hep-th/0004004];
%%CITATION = HEP-TH 0004004;%%
Phys.\ Rev.\ D {\bf 64} (2001) 024016 [hep-th/0101154];
%%CITATION = HEP-TH 0101154;%%
F.~C.~Adams, G.~L.~Kane, M.~Mbonye and M.~J.~Perry,
Int.\ J.\ Mod.\ Phys.\ A {\bf 16} (2001) 2399 [hep-ph/0009154];
%%CITATION = HEP-PH 0009154;%%
S.~B.~Giddings and S.~Thomas,
%``High energy colliders as black hole factories: The end of
%short  distance physics,''
Phys.\ Rev.\ D {\bf 65}, 056010 (2002) [hep-ph/0106219];
%%CITATION = HEP-PH 0106219;%%
S.~Dimopoulos and G.~Landsberg,
Phys.\ Rev.\ Lett.\  {\bf 87} (2001) 161602 [hep-ph/0106295];
%%CITATION = HEP-PH 0106295;%%
M.~B.~Voloshin, Phys.\ Lett.\ B {\bf 518} (2001) 137 [hep-ph/0107119];
%%CITATION = HEP-PH 0107119;%%
K.~m.~Cheung,
%``Black hole production and large extra dimensions,''
Phys.\ Rev.\ Lett.\  {\bf 88}, 221602 (2002) [hep-ph/0110163];
%%CITATION = HEP-PH 0110163;%%
%``Black hole, string ball, and p-brane production at hadronic supercolliders,''
hep-ph/0205033;
%%CITATION = HEP-PH 0205033;%%
%%
P.~Kanti and J.~March-Russell,
%``Calculable corrections to brane black hole decay. I: The scalar case,''
Phys.\ Rev.\ D {\bf 66}, 024023 (2002) [hep-ph/0203223];
%%CITATION = HEP-PH 0203223;%%
%%
A.~Chamblin and G.~C.~Nayak,
%``Black hole production at LHC: String balls and black holes from p p
% and  lead lead collisions,''
hep-ph/0206060.
%%CITATION = HEP-PH 0206060;%%

\bibitem{KT}
P.~Kanti and K.~Tamvakis,
%``Quest for localized 4-D black holes in brane worlds,''
Phys.\ Rev.\ D {\bf 65}, 084010 (2002) [hep-th/0110298].
%%CITATION = HEP-TH 0110298;%%

\bibitem{CK}
A.~G.~Cohen and D.~B.~Kaplan,
%``Solving the hierarchy problem with noncompact extra dimensions,''
Phys.\ Lett.\ B {\bf 470}, 52 (1999) [hep-th/9910132].
%%CITATION = HEP-TH 9910132;%%


%\cite{Kanti:2001vb}
\bibitem{Kanti:2001vb}
P.~Kanti, R.~Madden and K.~A.~Olive,
%``A 6-D brane world model,''
Phys.\ Rev.\ D {\bf 64}, 044021 (2001) [hep-th/0104177].
%%CITATION = HEP-TH 0104177;%%


\bibitem{Ola-1}
I.~Olasagasti and A.~Vilenkin,
%``Gravity of higher-dimensional global defects,''
Phys.\ Rev.\  {\bf D62}, 044014 (2000) [hep-th/0003300].
%%CITATION = HEP-TH 0003300;%%

\bibitem{Ola-2}
I.~Olasagasti,
%``Gravitating global defects: The gravitational field and
%compactification,''
Phys.\ Rev.\ D {\bf 63}, 124016 (2001) [hep-th/0101203].
%%CITATION = HEP-TH 0101203;%%


\bibitem{Kaloper}
N.~Kaloper, Phys.\ Rev.\ D {\bf 60}, 123506 (1999) [hep-th/9905210].
%%CITATION = HEP-TH 9905210;%%

\bibitem{Kim}
H.~B.~Kim and H.~D.~Kim,
Phys.\ Rev.\ D {\bf 61}, 064003  (2000) [hep-th/9909053].
%%CITATION = HEP-TH 9909053;%%


\bibitem{KR}
A.~Karch and L.~Randall,
JHEP {\bf 0105}, 008  (2001) [hep-th/0011156];
%%CITATION = HEP-TH 0011156;%%
%
I.~I.~Kogan, S.~Mouslopoulos and A.~Papazoglou,
%``A new bigravity model with exclusively positive branes,''
Phys.\ Lett.\ B {\bf 501}, 140 (2001) [hep-th/0011141];
%%CITATION = HEP-TH 0011141;%%
%
A.~Miemiec,
%``A power law for the lowest eigenvalue in localized massive gravity,''
Fortsch.\ Phys.\  {\bf 49}, 747 (2001) [hep-th/0011160];
%%CITATION = HEP-TH 0011160;%%
%
M.~D.~Schwartz,
%``The emergence of localized gravity,''
Phys.\ Lett.\ B {\bf 502}, 223 (2001) [hep-th/0011177];
%%CITATION = HEP-TH 0011177;%%
%
I.~Oda,
%``Localization of bulk fields on AdS(4) brane in AdS(5),''
Phys.\ Lett.\ B {\bf 508}, 96 (2001) [hep-th/0012013];
%%CITATION = HEP-TH 0012013;%%
%
%``Locally localized gravity models in higher dimensions,''
Phys.\ Rev.\ D {\bf 64}, 026002 (2001) [hep-th/0102147].
%%CITATION = HEP-TH 0102147;%%

\bibitem{HK}
T.~Hirayama and G.~Kang,
Phys.\ Rev.\ D {\bf 64} (2001) 064010 [hep-th/0104213].
%%CITATION = HEP-TH 0104213;%%

\bibitem{KOP1}
P.~Kanti, K.~A.~Olive and M.~Pospelov,
%``Static solutions for brane models with a bulk scalar field,''
Phys.\ Lett.\ B {\bf 481}, 386 (2000) [hep-ph/0002229].
%%CITATION = HEP-PH 0002229;%%

\bibitem{kkop}
P.~Kanti, I.~I.~Kogan, K.~A.~Olive and M.~Pospelov,
Phys.\ Lett.\ B {\bf 468}, 31  (1999) [hep-ph/9909481];
%%CITATION = HEP-PH 9909481;%%
Phys.\ Rev.\ D {\bf 61}, 106004  (2000) [hep-ph/9912266].
%%CITATION = HEP-PH 9912266;%%

\bibitem{self}
N.~Arkani-Hamed, S.~Dimopoulos, N.~Kaloper and R.~Sundrum,
Phys. Lett. B {\bf 480}, 193 (2000) [hep-th/0001197];
%%CITATION = HEP-TH 0001197;%%
S.~Kachru, M.~Schulz and E.~Silverstein, Phys. Rev. D {\bf 62}, 045021 (2000)
[hep-th/0001206]. %%CITATION = HEP-TH 00012066;%%

\bibitem{fine}
S.~F\"{o}rste, Z.~Lalak, S.~Lavignac and H.~P.~Nilles, Phys. Lett. B {\bf 481},
360 (2000) [hep-th/0002164]; %%CITATION = HEP-TH 0002164;%%
S.~P.~de~Alwis, Nucl. Phys. B {\bf 597} 263 (2001) [hep-th/0002174];
%%CITATION = HEP-TH 0002174;%%
C.~Csaki, J.~Erlich, C.~Grojean and T.~J.~Hollowood,
Nucl. Phys. B {\bf 584} 359 (2000) [hep-th/0004133].
%%CITATION = HEP-TH 0004133;%%

\bibitem{eff} C.~Charmousis, Class. Quant. Grav. {\bf 19}, 83 (2002)
[hep-th/0107126]; %%CITATION = HEP-TH 0107126;%%
S.~C.~Davis, JHEP {\bf 0203}, 058 (2002) [hep-ph/0111351].
%%CITATION = HEP-PH 0111351;%%

\bibitem{cline} J.~M.~Cline and H.~Firouzjahi,
Phys. Rev. D {\bf 65} 043501 (2002) [hep-th/0107198].



\end{thebibliography}
\end{document}